\documentclass[twocolumn,amsmath,amssymb,superscriptaddress,pra]{revtex4}

\usepackage{graphicx}
\usepackage{amsmath,amssymb,amsthm}

\newcommand{\beq}{\begin{equation}}
\newcommand{\eeq}{\end{equation}}
\newcommand{\beqa}{\begin{eqnarray}}
\newcommand{\eeqa}{\end{eqnarray}}
\newcommand{\e}{\mathrm{e}}
\newcommand{\w}{\omega}

\newcommand{\ket}[1]{\left| #1 \right\rangle}
\newcommand{\bra}[1]{\left\langle #1 \right|}

\newcommand{\av}[1]{\langle #1\rangle}
\newcommand{\braket}[2]{\langle #1 | #2\rangle}

\newcommand{\ketbra}[2]{\left|#1\right\rangle\hskip-1mm\left\langle #2\right|}

\newcommand{\Bav}{\langle B \rangle}
\newcommand{\txt}{\mathrm}
\newcommand{\average}[1]{\langle #1 \rangle}

\begin{document}

\title{Separation-dependent localization in a two-impurity spin-boson model}
\author{Dara P. S. McCutcheon} \email{dara.mccutcheon@ucl.ac.uk}
\affiliation{Department of Physics and Astronomy, University College London, Gower Street, London WC1E 6BT, United Kingdom}
\affiliation{London Centre for Nanotechnology, University College London}
\author{Ahsan Nazir} \email{ahsan.nazir@ucl.ac.uk}
\affiliation{Department of Physics and Astronomy, University College London, Gower Street, London WC1E 6BT, United Kingdom}
\author{Sougato Bose}
\affiliation{Department of Physics and Astronomy, University College London, Gower Street, London WC1E 6BT, United Kingdom}
\author{Andrew J. Fisher} 
\affiliation{Department of Physics and Astronomy, University College London, Gower Street, London WC1E 6BT, United Kingdom}
\affiliation{London Centre for Nanotechnology, University College London}

\date{\today}

\begin{abstract}
 
Using a variational approach we investigate the delocalized to localized crossover in the ground state of an Ohmic two-impurity spin-boson model, describing two otherwise non-interacting spins coupled to a common bosonic environment. 
We show that a competition between an environment-induced Ising spin interaction and externally applied fields leads to variations in the system-bath coupling strength, $\alpha_c$, at which the delocalized-localized crossover occurs. Specifically, the crossover regime lies between $\alpha_c=0.5$ and $\alpha_c=1$ depending upon the 
spin separation and the strength of the transverse tunneling field. 
This is in contrast to the analogous single spin case, for which the crossover occurs 
(in the scaling limit) at fixed $\alpha_c\approx1$. We also discuss links between the two-impurity spin-boson model and a dissipative two-spin transverse Ising model, showing that the latter possesses the same qualitative features as the Ising strength is varied. Finally, we show that signatures of the crossover may be observed in single impurity observables, as well as in the behaviour of the system-environment entanglement.

\end{abstract}


\maketitle

\section{Introduction}

Any real quantum system is surrounded by some uncontrollable environment, interactions with which 
generally lead to differing behaviour from that expected if the system were somehow isolated~\cite{b+p}. The spin-boson model~\cite{leggett87,weissbook} is a popular 
starting point for investigations into such dynamics, encapsulating the effects of quantum decoherence, dissipation 
and relaxation on the otherwise coherent spin evolution. Furthermore, the model exhibits non-trivial ground-state behaviour~\cite{leggett87, weissbook, lehur08, vojta2005, wong08, chin06, bulla2003}, displaying a zero-temperature (quantum) phase transition~\cite{vojta03, sachdev} as a function of system-bath coupling strength, attributed to zero-point (rather than thermal) fluctuations within the bath. Besides being of general theoretical interest, many physical systems in the solid state, and elsewhere, are well described by models of a spin-boson type~\cite{leggett87, weissbook, dube98, mahan, ramsay10, porras08, cedraschi2000, tong06, recati05, devoret97}. Specific experimentally relevant examples include large arrays of trapped ions~\cite{porras08}, the persistent current in a metal ring threaded by an Aharonov-Bohm flux~\cite{cedraschi2000, tong06}, and 
atomic dots coupled to a Bose-Einstein condensate bath~\cite{recati05}. These systems are of particular importance, since it is predicted that they show \emph{qualitative} and detectable changes in ground-state properties as a function of 
accessible external parameters. 

The spin-boson model considers a two-level quantum system, such as a spin-$1/2$ particle or a magnetic impurity, 
interacting with an (infinite) bath of harmonic oscillators representing the environment. The corresponding Hamiltonian is generally written in the form (for $\hbar=1$)
\beq
H=\frac{\epsilon}{2}\sigma_z-\frac{\Delta}{2}\sigma_x+\sigma_z\sum_{\bf k}(g_{\bf k} b_{\bf k}^{\dagger}+g_{\bf k}^*b_{\bf k})+\sum_{\bf k}\w_{\bf k}b_{\bf k}^{\dagger}b_{\bf k},
\label{eqnsinglespinbosonmodel}
\eeq
where $\epsilon$ is the energy bias between the system states, $\Delta$ is the (bare) tunneling strength, 
and $\sigma_i$ ($i=x,y,z$) is the usual $i$th-Pauli operator in a basis where $\sigma_z=\ketbra{0}{0}-\ketbra{1}{1}$.  
The bath is represented by the creation (anihilation) operators $b_{\bf k}^{\dagger}$ ($b_{\bf k}$) for each bath mode, with wave-vector 
${\bf k}$ and corresponding angular frequency $\w_{\bf k}$. The system-bath interaction is captured by the 
coupling constants $g_{\bf k}$. 

As is well known, the interaction of a quantum system with an environment of the type given in Eq.~(\ref{eqnsinglespinbosonmodel}) causes a renormalization of the bare system energy levels and, in particular, a suppression of any tunneling probability the system may possess~\cite{b+p, leggett87,weissbook}. For the spin-boson model, the system-bath interaction can be completely characterized 
by the spectral density $J(\omega)=\sum_{\bf k}|g_{\bf k}|^2\delta(\omega-\omega_{\bf k})$, which we shall take here to be of the paradigmatic Ohmic form $J(\omega)=(\alpha/2)\omega$ for $\omega<\omega_c$, where $\alpha$ is a dimensionless coupling strength and $\omega_c$ a high-frequency cutoff~\cite{b+p, leggett87,weissbook}. In this case, it has been found that above a certain critical system-bath coupling strength, $\alpha_c$, the tunneling probability is completely suppressed ($\Delta\rightarrow0$)~\cite{leggett87}. For small $\epsilon/\Delta$, as the parameter $\alpha$ is increased through $\alpha_c$, the ground state of the 
two-level system shows a crossover from being dominated by the tunneling term $(\Delta/2)\sigma_x$, and hence delocalised, to being dominated by the bias term $(\epsilon/2) \sigma_z$, and therefore localized in either $\ket{0}$ or $\ket{1}$~\cite{lehur08, lehur07}. 
At zero temperature and for $\epsilon=0$ this localization phenomenon has been identified as a Kosterlitz-Thouless quantum (rather than thermodynamic) phase transition~\cite{vojta03, sachdev}. Calculation of the Ohmic critical coupling strength in this regime has found $\alpha_c\approx1$ for small $\Delta/\omega_c$~\cite{leggett87}. 

In this paper, we shall investigate the delocalized-localized crossover in the ground state of a pair of non-interacting two-level systems in a common bath of harmonic oscillators, termed here the two-impurity spin-boson model. In particular, we elucidate how this crossover 
depends upon the separation between the impurities through a {\it bath-induced} inter-spin interaction. Aside from being a natural extension of the 
single impurity model, the two-impurity case represents perhaps the 
simplest dissipative model in which to explore the interplay of coherent system interactions and the dissipative influence of the bath. This has relevance, for example, in the field of quantum computation~\cite{n+c}, where the two-impurity model could be thought of as the basic unit of a dissipative spin chain~\cite{fisher1993, cincio2009, garnerone2009} 
or as two quantum bits in a dissipative register~\cite{sergi09, negele2008, dube98}. The model has also recently gained attention since it allows for the study of 
bath generated correlations and entanglement 
shared between the impurities~\cite{mccutcheon09, zell09, benatti03, solenov07,ficek08}. 

To perform our analysis we employ a variational 
technique originally developed by Silbey and Harris~\cite{silbey84}. The method consists of assuming a particular variational form of the ground-state wavefunction of the combined system and bath, 
and a subsequent optimization based upon a minimization of the associated free energy (or ground state energy at zero temperature). While this technique might be vulnerable to errors in certain limits, it has proven to be relatively robust when applied to single spin-boson systems described by Ohmic spectral densities~\cite{silbey84}. Furthermore, conclusions drawn from the 
method have also been verified by path integral~\cite{leggett87, bray1982}, flow equation~\cite{kehrein1996} and scaling techniques~\cite{stauber2006}, as well as by Bethe-ansatz~\cite{pono1993, costi99, lehur07}, Numerical Renormalization Group~\cite{lehur07, bulla2003}, Density Matrix Renormalization Group~\cite{wong08}, exact diagonalization~\cite{alvermann2009}, and Monte Carlo calculations~\cite{winter09}. Besides its relative simplicity, the variational technique is also attractive from the point of view of gaining insight into the form of the ground-state of the model, and how this varies through the delocalized to localized crossover. Furthermore, as we shall show below, it can be used to provide analytical calculations of bath-induced spin interaction terms and tunneling renormalizations inherent to the model. 

While it is known that a single un-biased two-level system immersed in a bosonic bath will enter its localized regime as the system-bath coupling strength $\alpha\rightarrow 1$ (for $\Delta/\omega_c\ll1$), it is not clear whether the 
same conclusion holds true for a pair of two-level systems immersed in a common bath. In this case, it is possible that the delocalized-localized crossover could occur at a different 
point due to an induced, bath-mediated interaction between the spins that alters the system energy structure~\cite{mccutcheon09, zell09, benatti03, solenov07}. For the model considered here, the mutual interaction with the bath induces an 
Ising-like coupling between the impurities with a strength that depends upon their separation. We find that for closely spaced impurities, corresponding to a strong Ising strength, the crossover region occurs around $\alpha_c=0.5$. As the spin separation is increased, the Ising strength decreases, and we find $\alpha_c \rightarrow 1$ as the impurity 
separation becomes infinite. 

The paper is organized as follows. In Section~\ref{secsinglespinbosonmodel} we review the application of the variational approach to the ground state of the standard single impurity spin-boson problem~\cite{silbey84}, demonstrating its validity in comparison to a Bethe-ansatz calculation~\cite{pono1993, lehur08}. In Section~\ref{sectwospinbosonmodel} we define the two-impurity model and apply the variational transformation in this situation, elucidating the origin of the bath-induced Ising interaction. In Section~\ref{variationalcalccrude} we employ an approximation on the induced Ising term that allows a straightforward identification of the delocalized-localized crossover region in a number of regimes, while in Section~\ref{secexactvariationalcalculation} we perform the full variational calculation without such a simplification. In Section~\ref{variationalgs} we study the variational ground state in more detail and propose signatures of the crossover behaviour, while in Section~\ref{secsummary} we summarise our results.

\section{Single impurity spin-boson model}
\label{secsinglespinbosonmodel}

Before we go on to study a pair of two-level systems in a common bath, it is instructive to apply the variational technique to the (single) spin-boson model, Eq.~(\ref{eqnsinglespinbosonmodel}), in an effort both to understand the variational method employed, and also to assess its validity. 

Let us start by considering the ground state of the Hamiltonian with $\epsilon=0$: 
\beq
H=-\frac{\Delta}{2}\sigma_x+\sigma_z\sum_{\bf k}(g_{\bf k} b_{\bf k}^{\dagger}+g_{\bf k}^*b_{\bf k})+\sum_{\bf k}\w_{\bf k}b_{\bf k}^{\dagger}b_{\bf k}.
\label{eqnsinglespinbosonmodelnobias}
\eeq
In the limit $g_{\bf k} \rightarrow 0$ with $\Delta \neq0$, the spin is entirely decoupled from the bath and its ground state will be the $\sigma_x$ eigenstate 
$(1/\sqrt{2})(\ket{0}+\ket{1})$. The state of the bath will be some superposition of number states (eigenstates of 
$b_{\bf k}^{\dagger}b_{\bf k}$) that depends upon the temperature. In the opposite limit, 
$\Delta \rightarrow 0$ with $g_{\bf k} \neq 0$, the system-bath interaction now dominates and the oscillators constituting 
the bath will be displaced from their equilibrium positions to minimize the corresponding interaction energy. We may write 
the ground state of the combined system-plus-bath in this case as 
\beq
\ket{\Psi}=\frac{1}{\sqrt{2}}\left(\prod_{\bf k}D\Bigl(\frac{g_{\bf k}}{\w_{\bf k}}\Bigr)\ket{B_0}\ket{1}+\prod_{\bf k} D\Bigl(-\frac{g_{\bf k}}{\w_{\bf k}}\Bigr)\ket{B_0}\ket{0}\right),
\label{eqnpolarongroundstate}
\eeq
where $\ket{B_0}$ is the ground state of the bath for vanishing system-bath coupling, and we have defined the displacement operators~\cite{glauber63}
\beq
D\Bigl(\pm\frac{ g_{\bf k}}{\w_{\bf k}}\Bigr)=\txt{exp}\bigg[\pm\bigg(\Bigl(\frac{g_{\bf k}}{\w_{\bf k}}\Bigr)b_{\bf k}^{\dagger}- 
\Bigl(\frac{g_{\bf k}}{\w_{\bf k}}\Bigr)^*b_{\bf k}\bigg)\bigg].
\label{fulldisplace}
\eeq

In the general case, when neither limit is met, the spin-boson Hamiltonian is not straightforwardly diagonalizable. Note, however, that in the state described by Eq.~(\ref{eqnpolarongroundstate}) each oscillator is 
displaced by an amount determined by the the ratio $g_{\bf k}/\w_{\bf k}$, and that as $g_{\bf k}\rightarrow0$ the correct uncoupled ground state is recovered. The variational 
theory thus assumes that the ground state of the spin-boson Hamiltonian for non-zero $g_{\bf k}$ and $\Delta$ is always of the form of Eq.~(\ref{eqnpolarongroundstate}), but allows for the possibility that the amount a given mode is displaced may have a more complicated dependence on the Hamiltonian parameters.

With these considerations in mind, we now reintroduce the energy bias between the spin states and proceed by writing down the total Hamiltonian, Eq.~(\ref{eqnsinglespinbosonmodel}), in a basis 
$\{\ket{B_-}\ket{0},\ket{B_+}\ket{1}\}$, where $\ket{B_{\pm}}=\prod_{\bf k} D(\pm f_{\bf k}/\w_{\bf k})\ket{B_0}$, and we have introduced the as yet to be determined variational parameters $f_{\bf k}$. These will be found by minimising the free energy of the total system~\cite{silbey84}. At zero temperature, we obtain
\beq
H=\frac{\epsilon}{2}\tilde{\sigma}_z-\frac{\Delta_r}{2}\tilde{\sigma}_x+R,
\eeq
where in the new basis $\tilde{\sigma}_z=\ket{B_-}\ket{0}\bra{0}\bra{B_+}-\ket{B_+}\ket{1}\bra{1}\bra{B_-}$ and 
\beq
R=\sum_{\bf k}\w_{\bf k}^{-1}f_{\bf k}(f_{\bf k}-2g_{\bf k}). 
\label{eqn:R}
\eeq
Importantly, the tunneling matrix element has now been renormalized due to the system-bath interaction: $\Delta_r=\Delta \av{B}$, where
\beq\label{Bavezerotemp}
\av{B}=\braket{B_{\pm}}{B_{\pm}}=\txt{exp}\bigg[-2\sum_{\bf k}(f_{\bf k}/\w_{\bf k})^2\bigg].
\eeq
Diagonalization of $H$ in the transformed basis then gives a ground state energy of $\lambda_0=\frac{1}{2}(2 R - \eta)$, where 
$\eta=\sqrt{\epsilon^2+\Delta_r^2}$, and the corresponding ground state
\beq
\ket{\phi_0}=n_0\left(\frac{\eta-\epsilon}{\Delta_r}\ket{B_-}\ket{0}+\ket{B_+}\ket{1}\right),
\eeq
where $n_0=((\eta-\epsilon)^2/\Delta_r^2+1)^{-1/2}$ is a normalisation factor. 

The task now is to find the variational parameters $f_{\bf k}$, which in turn will allow us to evaluate the renormalized tunneling strength $\Delta_r$, and hence obtain the approximate ground state. 
To do so, we naturally impose the condition that the ground-state energy, $\lambda_0$, 
should be minimised. This leads straightforwardly to
\beq
f_{\bf k}=g_{\bf k}\left(1+\frac{\Delta_r^2}{\w_{\bf k} \eta}\right)^{-1},
\label{eqnfkminwithepsilon}
\eeq
and our expression for the renormalized tunneling strength becomes
\beq\label{deltarzerotemp}
\Delta_r=\Delta\exp{\left[-2\sum_{\bf k} \frac{g_{\bf k}^2}{(\w_{\bf k}+\Delta_r^2/\eta)^2}\right]}.
\eeq
We now take the continuum limit to convert the summation over ${\bf k}$ into an integral with respect to $\w$, and recall the definition of the (Ohmic) system-bath spectral density, $J(\w)=\sum_{\bf k} |g_{\bf k}|^2\delta(\w-\w_{\bf k})=(\alpha/2)\omega$. With these replacements, we find 
\beq
\Delta_r=\Delta \exp{\left[-\alpha\int_0^{\w_c} \frac{\w\,d\w}{(\w+\Delta_r^2/\eta)^2}\right]}.
\label{detarint}
\eeq
Note that had we written the original Hamiltonian in a basis defined with the displacement operators of Eq.~(\ref{fulldisplace}), i.e. functions of $g_{\bf k}$ rather than $f_{\bf k}$, the integral in Eq.~(\ref{detarint}) would 
suffer from an infra-red divergence, and we would conclude (incorrectly) that $\Delta_r=0$ for all values of $\alpha$. 

In the present case, the integration can be performed 
straightforwardly and leads to the following equation which one must solve self-consistently for $\Delta_r$:
\beq
\Delta_r\bigg(\frac{\Delta_r^2}{\Delta_r^2+\w_c \eta}\bigg)^{-\alpha}\txt{exp}\bigg[\frac{-\alpha\,\w_c \eta}{\Delta_r^2+\w_c \eta}\bigg]=\Delta,
\label{eqndeltarselfwithepsilon}
\eeq
where $\Delta_r$ takes on values between $\Delta$ and $0$ as $\alpha$ is increased from zero. For $\Delta/\omega_c\ll1$ and $\epsilon=0$, Eq.~(\ref{eqndeltarselfwithepsilon}) gives the well-known behaviour $\Delta_r\sim\Delta(\Delta/\omega_c)^{\alpha/(1-\alpha)}$~\cite{leggett87,weissbook,silbey84}.

\begin{figure}[!t]
\begin{center}
\includegraphics[width=0.45\textwidth]{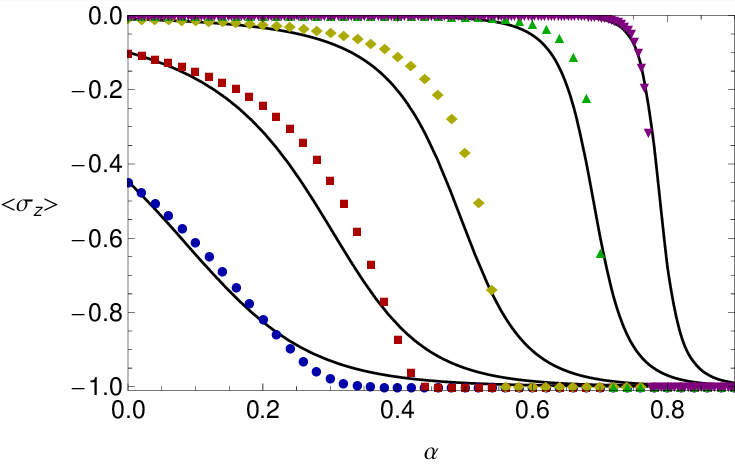}
\caption{(Color online) Expectation value of $\sigma_z$ for the variationally determined spin-boson ground state (plot points) and using Bethe-ansatz techniques (solid lines) 
as a function of $\alpha$, plotted for various values of $\epsilon$ (in units of $\omega_c$). Blue circles: $\epsilon=0.005$; 
red squares: $\epsilon=10^{-3}$; yellow diamonds: $\epsilon=10^{-4}$; green upright triangles: $\epsilon=10^{-6}$; and purple inverted triangles: $\epsilon=10^{-8}$. In all cases $\Delta/\w_c=0.01$.}
\label{figsigmaz}
\end{center}
\end{figure}

In order to assess the validity of the variational technique, in Fig.~\ref{figsigmaz} we plot the ground state magnetisation, $\av{\sigma_z}=\bra{\phi_0}\sigma_z\ket{\phi_0}=-\epsilon/\eta$, as a function of $\alpha$ for various values of $\epsilon$, where we have set 
$\Delta/\w_c=0.01$ and $\Delta_r$ has been found by numerically solving Eq.~(\ref{eqndeltarselfwithepsilon}). Shown also are the corresponding plots generated by mapping the spin-boson model to the Kondo model and using Bethe-ansatz solution techniques, details of which can 
be found in Refs.~\cite{pono1993, lehur08, costi99}. For all values of $\epsilon$ the methods show good qualitative agreement. 
Most importantly, the variational calculation correctly identifies the region of $\alpha$ over which the ground state becomes dominated by the 
bias (localization) rather than the tunneling (delocalization), though it should be noted that $\av{\sigma_z}$ reaches its minimum value ($\av{\sigma_z}\rightarrow-1$ as $\Delta_r\rightarrow 0$) somewhat more sharply than in the Bethe-ansatz calculations. We can therefore be confident that the variational method does capture the localization crossover in the ground state behaviour in which we are interested.

\section{Two-impurity spin-boson model}
\label{sectwospinbosonmodel}

We now return our attention to the main subject of this work, determining the ground state behaviour of a pair of impurity spins interacting with a common bosonic bath. Since we expect the bath to mediate a separation-dependent coherent interaction between the spins~\cite{vorrath05, dube98, mccutcheon09, benatti03, solenov07} we make their spatial separation explicit by placing them at positions ${\bf r}_1$ and ${\bf r}_2$. 
The total Hamiltonian we consider is then given by 
\begin{eqnarray}\label{2spinboson}
H&{}={}&-\frac{\Delta}{2}(\sigma_x^1+\sigma_x^2)+\sum_{\bf k}\w_{\bf k} b_{\bf k}^{\dagger}b_{\bf k}\nonumber\\
&&\:{+}\sum_n\sigma_z^n\sum_{\bf k} g_{\bf k} \bigl(b_{\bf k}^{\dagger}\e^{i {\bf k}\cdot {\bf r}_n}+b_{\bf k}\e^{-i{\bf k}\cdot{\bf r}_n}\bigr),
\end{eqnarray}
where $\sigma^n_i$ ($n=1,2$; $i=x,y,z$) is now
the usual $i$th-Pauli operator acting on the relevant spin, and we have assumed that the system-bath coupling constants for each spin differ only in a position-dependent phase factor.
For simplicity, we now limit our analysis to the case in which there is no bias on either spin.

We proceed in a slightly different manner to Section~\ref{secsinglespinbosonmodel} and apply a unitary transformation to the Hamiltonian which displaces each oscillator by an amount parameterized 
by the variational parameter $f_{\bf k}$. We note, however, that at zero temperature this 
procedure gives the same ground state as one would obtain following the method of the previous section. 
The transformed Hamiltonian is written
$\tilde{H}=\e^{S_1+S_2}H\e^{-(S_1+S_2)}=\tilde{H}_S+\tilde{H}_B+\tilde{H}_I$ with
\beq
\e^{\pm S_n}=\txt{exp}\left[\pm \sigma_z^n\sum_{\bf k}\alpha_{\bf k}\bigl(b_{\bf k}^{\dagger}\e^{i{\bf k}\cdot{\bf r}_n}-b_{\bf k}\e^{-i{\bf k}\cdot{\bf r}_n}\bigr)\right],
\eeq
where $\alpha_{\bf k}=f_{\bf k}/\w_{\bf k}$ is assumed real. The transformation is aided by the observation that, provided the dispersion relation of the bath is isotropic and 
the variational parameters satisfy $f_{\bf{k}}=f_{-\bf{k}}$, 
the commutator
\beq
[S_1,S_2]=2 i \sigma_z^1\sigma_z^2\sum_{\bf k}\alpha_{\bf k}^2\sin\left({\bf k}\cdot({\bf r}_1-{\bf r}_2)\right)
\label{eqn_s_commutator}
\eeq
vanishes once the summation has been performed, regardless of the dimensionality or frequency spectrum of the system-bath interaction. 

The variational technique now relies on a careful choice of $\tilde{H}_S$, $\tilde{H}_B$ and $\tilde{H}_I$ from the 
various terms available after the transformation. We define the new unperturbed Hamiltonian as $\tilde{H}_0=\tilde{H}_S+\tilde{H}_B$, with 
\beq
\tilde{H}_S=-\frac{\Delta_r}{2}\bigl(\sigma_x^1+\sigma_x^2\bigr)-2X\sigma_z^1\sigma_z^2
+2\sum_{\bf k}f_{\bf k}(f_{\bf k}-2g_{\bf k})/\omega_{\bf k},
\label{eqntildeH_0}
\eeq
and $\tilde{H}_B=H_B$. Here, $\Delta_r$ is now determined by the finite temperature generalization of Eq.~(\ref{Bavezerotemp}), and is given by
\beq
\Delta_r=\Delta \Bav=\Delta\exp{\bigg[-2\sum_{\bf k}\alpha_{\bf k}^2\txt{coth}(\beta\w_{\bf k}/2)\bigg]},
\label{eqndeltardefinition}
\eeq
where the inverse temperature is $\beta=1/k_BT$, while the form of $\tilde{H}_S$ has been chosen such that the expectation value of $\tilde{H}_I$ with respect to $\tilde{H}_{0}$ vanishes. We shall see that this significantly simplifies the determination of the $\{f_{\bf k}\}$ below.

There then remain two forms of system-bath interaction, $\tilde{H}_I=\tilde{H}_z+\tilde{H}_{\perp}$, where
\beq
\tilde{H}_z=\sum_n\left(\sigma_z^n\sum_{\bf k}(g_{\bf k}-f_{\bf k})\Bigl(b^{\dagger}_{\bf k}\e^{i{\bf k}\cdot {\bf r}_n}+b_{\bf k}\e^{-i{\bf k}\cdot {\bf r}_n}\Bigr)\right),
\label{eqntildeH_z}
\eeq
and
\beq
\tilde{H}_{\perp}=-\frac{\Delta}{2}\sum_n\bigg((B_+^{(n)}-\Bav)\sigma_+^n+(B_-^{(n)}-\Bav)\sigma_-^n\bigg)
\label{eqntildeH_pm},
\eeq
with $\sigma_{\pm}^n=(1/2)(\sigma_x^n\pm i\sigma_y^n)$, and bath operators again given by products of the displacement operators: 
\beq
B_{\pm}^{(n)}=\exp{\left[\pm2\sum_{\bf k}\alpha_{\bf k}(b_{\bf k}^{\dagger}\e^{i{\bf k}\cdot {\bf r}_n}-b_{\bf k}\e^{-i{\bf k}\cdot {\bf r}_n})\right]}.
\eeq
Note that if we assume the bath to be in thermal equilibrium, these four bath operators all have same expectation value with respect to $\tilde{H}_0$: $\smash{\Bav=\langle B^{(n)}_{\pm}\rangle_{\tilde{H}_0}}=\exp{\left[-2\sum_{\bf k}\alpha_{\bf k}^2\txt{coth}(\beta\w_{\bf k}/2)\right]}$.

The unperturbed Hamiltonian, $\tilde{H}_0$, has two important features. Firstly, the tunneling strength, $\Delta_r$, has been renormalized. Secondly, the two spins are now coupled via a bath mediated, separation-dependent, Ising-like interaction, with a strength
\beq
X=\sum_{\bf k}\w_{\bf k}^{-1}f_{\bf k}(2g_{\bf k}-f_{\bf k})\cos({\bf k}\cdot({\bf r}_1-{\bf r}_2)).
\label{eqnXdefinition}
\eeq
Evaluation of both $\Delta_r$ and $X$ requires knowledge of the set of variational parameters $\{f_{\bf k}\}$. The 
variational procedure determines these by free energy minimization arguments. However, before we continue the 
analysis, we outline a significant simplification which can be made.

\section{Variational calculation}
\label{variationalcalccrude}

\subsection{Crude Ising approximation}
\label{seccrudeising}

The variational parameters $\{f_{\bf k}\}$, appearing in both $\Delta_r$ [Eq.~(\ref{eqndeltardefinition})] and 
in the induced Ising strength $X$ [Eq.~(\ref{eqnXdefinition})], were introduced to overcome an infra-red divergence in $\Delta_r$ that occurs for an Ohmic spectral density when applying a polaron transformation to our Hamiltonian, since it fully displaces the bath modes as in Eq.~(\ref{eqnpolarongroundstate})~\cite{silbey84}. As mentioned previously, this divergence would lead to a complete suppression of the tunneling probability, $\Delta_r\rightarrow0$, and can be seen by making the replacement $f_{\bf k}\rightarrow g_{\bf k}$ in Eq.~(\ref{eqndeltardefinition}), and using $J(\omega)=(\alpha/2)\omega$.

\begin{figure}[!t]
\begin{center}
\includegraphics[width=0.45\textwidth]{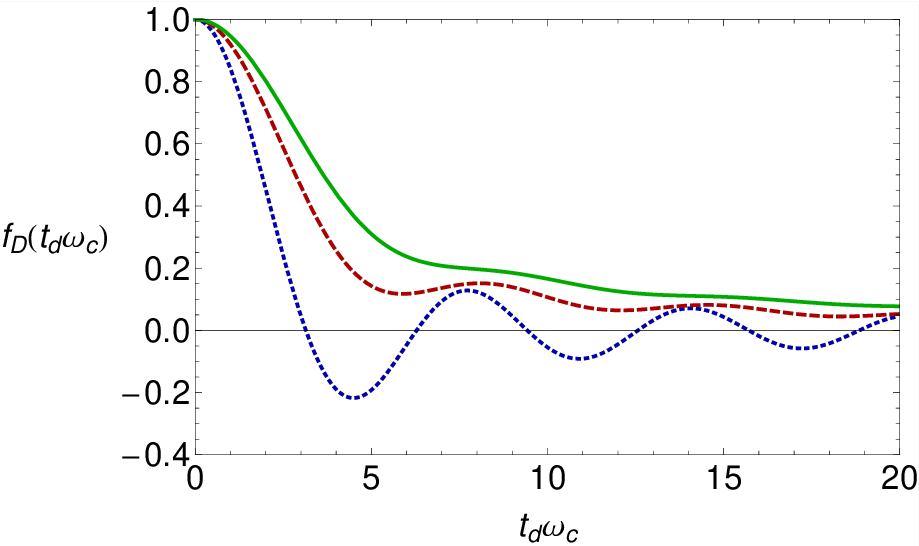}
\caption{(Color online) Measure of the correlation between the bath-induced fluctuations experienced at each impurity spin, plotted as a function of the scaled impurity separation $t_d\omega_c=|{\bf r}_1-{\bf r}_2|\omega_c/c$. The different curves correspond to system bath coupling in one dimension (blue dotted curve), two dimensions (dashed red curve), and three dimensions (solid green curve). In all cases the correlation is maximised for zero separation ($f_D =1$, complete correlation) and tends to zero as the separation goes to infinity ($f_D=0$, no correlation).}
\label{f_Dplots}
\end{center}
\end{figure}

However, we may make the replacement $f_{\bf k}\rightarrow g_{\bf k}$ in the definition of the Ising strength [Eq.~(\ref{eqnXdefinition})] and 
find that it suffers from no such divergence. Therefore, to some level of approximation at least, we can make this replacement (in Eq.~(\ref{eqnXdefinition}) only) and evaluate 
$X$ outside the variational calculation. We know that in the limit that the coupling of the system to the bath completely dominates, the oscillators 
are fully displaced, i.e. $f_{\bf k} \rightarrow g_{\bf k}$ anyway. Hence, we can identify this replacement as a kind of strong coupling approximation on $X$, as will be discussed in more detail in Section~\ref{isingcomp}.
We shall refer to the Ising term evaluated within this approximation, and to the approximation itself, as ``crude'' since it does not take into account deviations of $f_{\bf k}$ from $g_{\bf k}$ in $X$. 

Assuming a linear dispersion relation $|{\bf k}|=\w/c$, where $c$ is the excitation speed, we find 
\beq
X_C=\frac{\alpha \w_c}{2} f_D(t_d \w_c),
\label{eqncrudeXallD}
\eeq
where the subscript $C$ indicates a crude value. The impurity distance dependence enters through $t_d=|{\bf r}_1-{\bf r}_2|/c$, which is the time bosonic excitations take to travel between the spins, and determines the value of the function $f_D(x)$ ($D=1,2,3$), which is a measure of the (separation-dependent) correlation between the bath influences seen at each spin, and is therefore dependent on the dimensionality $D$ of the system-bath interaction. We find $f_1(x)=\txt{sinc}(x)$ in one dimension, $f_2(x)={_1F_2}\left(\{1/2\},\{1,3/2\},-x^2/4\right)$ is a generalized hypergeometic function in two dimensions, and $f_3(x)=\txt{Si}(x)/x$ in three dimensions, $\txt{Si}(x)=\int_0^x(\sin{t}/t)dt$ being the sine integral function. 
As shown in Fig.~\ref{f_Dplots}, in all cases $f_D(x)$ has a maximum value of $f_D(0)=1$, and in two and three dimensions has a minimum value $f_D(\infty)=0$. Additionally, 
in one dimension $f_1(x)$ displays decaying oscillations, becoming zero whenever $t_d=n \pi \w_c$, for $n=1,2,3,\dots$. 

Note that, ignoring any spatial correlations in $\smash{B_{\pm}^{(1)}}$ and $\smash{B_{\pm}^{(2)}}$, our transformed Hamiltonian $\tilde{H}$ with the replacement $X\rightarrow X_C$ now has exactly the same form as that which would be obtained if we transformed a Hamiltonian describing two spins in \emph{separate} baths, each subject to a transverse field of strength $\Delta$, and coupled via a ferromagnetic Ising field of strength $2 X_C$~\cite{werner05,orth08}. That is, had we transformed the 
Hamiltonian
\begin{eqnarray}
H_{TI}&{}={}&-\frac{\Delta}{2}(\sigma_x^1+\sigma_x^2)-2X_C\sigma_z^1\sigma_z^2\nonumber\\
&&\:{+}\sigma_z^1\sum_{\bf k}(g_{\bf k} b_{\bf k}^{\dagger}+g_{\bf k}^*b_{\bf k})+\sigma_z^2\sum_{\bf k}(g_{\bf k} a_{\bf k}^{\dagger}+g_{\bf k}^*a_{\bf k})\nonumber\\
&&\:{+}\sum_{\bf k}\w_{\bf k} b_{\bf k}^{\dagger}b_{\bf k}+\sum_{\bf k} \nu_{\bf k} a_{\bf k}^{\dagger}a_{\bf k},
\end{eqnarray}
where we have introduced a second bath which couples only to the second spin and is described 
by creation (annihilation) operators $a_{\bf k}^{\dagger}$ ($a_{\bf k}$), with corresponding frequencies $\nu_{\bf k}$. 

\subsection{Free energy minimization}
\label{secvariationalcalculation}

Precisely as in the single-spin case, our task is now to find the set of variational parameters $\{f_{\bf k}\}$, which will then allow us to find the renormalised tunneling strength $\Delta_r$. If, 
for a given $\alpha$, we find that $\Delta_r\rightarrow 0$ (i.e. $f_{\bf k}\rightarrow g_{\bf k}$), the system will be dominated by the induced Ising interaction, forming a ferromagnetic or antiferromagnetic pair. The spins will be unable tunnel between their 
states $\ket{0}$ and $\ket{1}$ and will be said to be in a localised regime. On the other hand, if $\Delta_r\neq 0$, the 
tunneling probability remains finite and the spins are delocalized. We expect that as $\alpha \rightarrow \alpha_c$, $\Delta_r \rightarrow 0$, and we enter a regime in which the renormalised tunneling has a negligible 
influence on the ground state.

To find the set $\{f_{\bf k}\}$ we follow Refs.~\cite{chin06} and~\cite{silbey84} and compute the Bogoliubov-Feynman upper bound on the free 
energy of the total system-plus-bath, $A_B$, which is related to the true free energy, $A$, via $A_B \geq A$~\cite{fisher}, where
\beq
A_B=-\beta^{-1}\txt{ln}\txt{Tr}\{\txt{exp}[-\beta \tilde{H}_0]\}+\average{\tilde{H}_I}_{\tilde{H}_0}+\mathcal{O}\bigl(\average{\tilde{H}_I^2}_{\tilde{H}_0}\bigr).
\label{eqnfeynmanfreeenergy}
\eeq
We have constructed our perturbation terms $\tilde{H}_I$ and system Hamiltonian $\tilde{H}_0$ such that $\average{\tilde{H}_I}_{\tilde{H}_0}=0$ by definition. We shall assume that terms of order $\average{\tilde{H}_I^2}_{\tilde{H}_0}$ are small, 
as shown in Ref.~\cite{chin06}, and approximate the free energy using only the first term of Eq.~(\ref{eqnfeynmanfreeenergy}). Neglecting the free energy of the bath, since it does not depend 
on the variational parameters, we find
\beq
\begin{split}
A_B\approx &\, 2\sum_{\bf k}\w_{\bf k}^{-1}f_{\bf k}(f_{\bf k}-2g_{\bf k})\\
&-\beta^{-1}\txt{ln}\Bigl[2\bigl(\txt{cosh}(2\beta X_C)+\txt{cosh}(\beta E_C)\bigr)\Bigr],
\end{split}
\label{eqnfreeenergywithX}
\eeq
where $E_C=\sqrt{4 X_C^2+\Delta_r^2}$. Minimizing $A_B$ with respect to the variational parameters yields the choice
\beq
f_{\bf k}=g_{\bf k}\bigg(1+\frac{\Delta_r^2}{\w_{\bf k} E_C}\bigg(\frac{\txt{sinh}(\beta E_C)\,\txt{coth}(\beta \w_{\bf k}/2)}{\txt{cosh}(2\beta X_C)+\txt{cosh}(\beta E_C)}\bigg)\bigg)^{-1}.
\label{eqnfkmintwospins}
\eeq
As we are interested here in the ground state (zero temperature) behaviour of the system we take 
the limit $\beta\rightarrow \infty$ to find
\beq
f_{\bf k}=g_{\bf k}\bigg(1+\frac{\Delta_r^2}{\w_{\bf k} E_C}\bigg)^{-1}.
\label{eqnfkmintwospinszt}
\eeq

Having found the optimal choice for each $f_{\bf k}$ in Eq.~(\ref{eqnfkmintwospinszt}), we can now insert this into our expression for the renormalised tunneling strength, Eq.~(\ref{eqndeltardefinition}). Taking the continuum limit 
and using the same form of Ohmic spectral density as before, we obtain the following self-consistent equation
\beq
\Delta_r\bigg(\frac{\Delta_r^2}{\Delta_r^2+\w_c E_C}\bigg)^{-\alpha}\txt{exp}\bigg[\frac{-\alpha\,\w_c E_C}{\Delta_r^2+\w_cE_C}\bigg]=\Delta.
\label{eqndeltarselfwithX}
\eeq
Note that with the replacement  $E_C\rightarrow \eta$ (or $2X_C\rightarrow \epsilon$) this equation is identical to Eq.~(\ref{eqndeltarselfwithepsilon}) derived in Section~\ref{secsinglespinbosonmodel} when considering a single spin with finite bias. This stems from the observation that, from the point of view of one of the spins, its Ising-like coupling to the other spin can be thought of as providing an effective energy difference between its $\sigma_z$ eigenstates.

The solutions of Eq.~(\ref{eqndeltarselfwithX}) give values of $\Delta_r$ that correspond to stationary points of the free energy approximation $A_B$. 
That a given solution exists does not necessarily mean that it is appropriate to assume that the system will adopt this value. Rather, we assume that (within our approximate treatment) the system will adopt the value of $\Delta_r$ that gives the lowest $A_B$~\cite{chen08}. To see which solution for $\Delta_r$ will be favored, 
we compute the free energy at zero temperature using the variational parameters we have just derived in Eq.~(\ref{eqnfkmintwospinszt}): 
\beq
A_B\approx-E_C-\alpha E_C \bigg(\frac{\Delta_r^2}{\w_c^2}+\frac{E_C}{\w_c}\bigg)^{-1}.
\eeq
Since we are working in the limit $\Delta/\w_c \ll 1$, it must also be true that $(\Delta_r/\w_c)^2 \ll E_C/\w_c$, 
regardless of the value of $X_C$, and we can further approximate  
\beq
A_B\approx-\alpha \w_c-E_C\bigg(1-\alpha\frac{\Delta_r^2}{E_C^2}\bigg).
\label{eqnA_Bapprox}
\eeq
The system will adopt which ever value of $\Delta_r$ makes the second term in Eq.~(\ref{eqnA_Bapprox}) most negative, i.e. that value which 
most strongly satisfies the condition $4X_C^2>\Delta_r^2(\alpha-1)$. For $\alpha<1$ it is clear that this will correspond to the greatest positive value of $\Delta_r$. 
Therefore, where multiple solutions to Eq.~(\ref{eqndeltarselfwithX}) exist, for $\alpha<1$ we should choose the largest value of $\Delta_r$.

\subsection{Separation-dependent localization}

In general, solving Eq.~(\ref{eqndeltarselfwithX}) for $\Delta_r$ analytically is not possible and it must be solved for numerically instead. However, to 
begin with, note that $\Delta_r=0$ is always a solution, regardless of the value of $\alpha$ or $X_C$. Now, we can look for other analytical solutions in certain limits. 
Perhaps the simplest of these is the limit $X_C \rightarrow 0$, corresponding to two infinitely separated spins in a common bath or two 
uncoupled spins in separate baths. From either interpretation, we should recover the well-known single spin-boson results. Setting 
$X_C=0$ in Eq.~(\ref{eqndeltarselfwithX}) gives
\beq
\Delta_r\left(\frac{\Delta_r}{\Delta_r+\w_c}\right)^{-\alpha}\txt{exp}\left[\frac{-\alpha\w_c}{\Delta_r+\w_c}\right]=\Delta,
\label{eqndeltarself}
\eeq
which in the limit $\Delta/\w_c \ll 1$ ($\Delta_r \leq \Delta$) gives the well-established form~\cite{silbey84, leggett87, weissbook} 
\beq
\Delta_r\approx\Delta\bigg(\frac{\e \Delta}{\w_c}\bigg)^{\alpha/(1-\alpha)}.
\label{eqn_standard_deltar}
\eeq
Hence, for $X_C=0$ and $\Delta/\w_c\ll 1$, the renormalised tunneling strength smoothly reaches zero as $\alpha \rightarrow 1$, and 
we predict that the critical coupling strength separating the delocalized and localized phases is given by $\alpha_c(X_C=0)=1$, precisely as in the single spin-boson case.

Let us now consider the opposite limit, $X_C/\Delta\gg 1$, corresponding either to closely spaced spins in a common bath with intermediate or strong dissipation 
(so that $X_C$ is large), or two spins in separate baths coupled via a relatively strong Ising interaction. Since $\Delta_r\leq\Delta$, we may also assume $X_C/\Delta_r \gg 1$. 
Setting $E_C\approx 2X_C$ and neglecting $\Delta_r$ in the denominators in both bracketed factors in Eq.~(\ref{eqndeltarselfwithX}), we find
\beq
\Delta_r\approx\Delta \bigg(\frac{\e \Delta^2}{2 \w_c X_C}\bigg)^{\alpha/(1-2 \alpha)}.
\label{eqnbigXdeltar}
\eeq
Within the limits this expression has been derived, the bracketed factor is small and we observe  
that $\Delta_r\rightarrow 0$ as $\alpha \rightarrow 0.5$. We conclude that for an Ising strength 
$X_C$ much larger than the bare tunneling strength, the critical system-bath coupling strength is no longer given by the single spin-boson value ($\alpha_c\approx1$), but instead by $\alpha_c(X_C/\Delta \gg1)\approx0.5$. This can also be seen by expanding our expression for the minimised free energy, Eq.~(\ref{eqnA_Bapprox}), to
lowest order in $\Delta_r/X_C$, which gives
\beq
A_B\approx -\alpha \w_c-2X_C\bigg(1+\frac{\Delta_r^2}{8X_C^2}(1-2\alpha)\bigg).
\label{eqnA_Bapprox2}
\eeq
It is clear from this expression that in the limit $X_C\gg\Delta_r$ a finite $\Delta_r$ will be favoured only if $\alpha<0.5$.

\begin{figure}[!t]
\begin{center}
\includegraphics[width=0.45\textwidth]{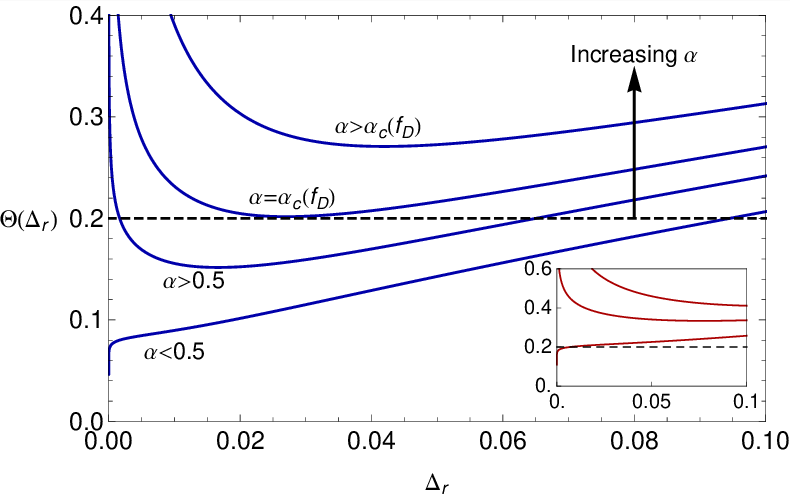}
\caption{(Color online) The renormalised tunneling strength is the solution to $\Theta(\Delta_r)=\Delta$ (see Eq.~(\ref{eqndeltarselfwithX})). Main: Here we plot $\Theta(\Delta_r)$ for various values of $\alpha$ and for a fixed spin separation 
corresponding to $f_D(t_d\w_c)=0.05$. Setting $\w_c=1$ then gives $X_C=0.025 \alpha$. We see that as $\alpha$ is increased the solution for $\Delta_r$ decreases until a point at which the curve $\Theta(\Delta_r)$ 
just touches the line $\Delta=0.2$. After this point, the only solution is $\Delta_r=0$. Inset: Here we plot the same curves but with $f_D(t_d\w_c)=0.3$. In this case $f_D$ (and hence $X_C$) is so large that $\alpha_c=0.5$, since above this value $\Theta(\Delta_r)$ never crosses $\Delta$.}
\label{figtheta}
\end{center}
\end{figure}

When neither of these conditions are met, i.e. when $X_C/\Delta \sim 1$, Eq.~(\ref{eqndeltarselfwithX}) is best studied graphically. To 
do so, we define the left hand side of Eq.~(\ref{eqndeltarselfwithX}) as a function $\Theta(\Delta_r)$. Any points at which
$\Theta(\Delta_r)$ crosses the line $\Delta$ will then give non-zero solutions for $\Delta_r$. In the main part of Fig.~\ref{figtheta} we plot $\Theta(\Delta_r)$ for a fixed spin separation, corresponding to a value of $f_D(t_d\w_c)=0.05$, and for various values of 
$\alpha$. Also shown is the dashed line at $0.2$, which represents the value of the bare tunneling strength $\Delta$ (in units of $\omega_c$) taken here. The first feature to notice is that the curve shows a 
dramatic change in behavior as the coupling strength $\alpha$ moves through the value $0.5$, developing a minimum in the first quadrant for $\alpha>0.5$. Therefore, when $\alpha<0.5$, we can always expect a single finite solution for $\Delta_r$. On the other hand, when $\alpha>0.5$, depending on the specific values of the 
ratio $X_C/\Delta$ and $\alpha$, the curve $\Theta(\Delta_r)$ may not cross the line $\Delta$ at all, just touch it, or dip low enough to cross it twice. 

In the main part of Fig.~\ref{figtheta} the ratio $X_C/\Delta$ is small enough such that $\Theta(\Delta_r)$ does dip below $\Delta$ for $\alpha\geq 0.5$, and the critical coupling 
strength is then given when $\Theta(\Delta_r)$ just touches the line $\Delta$. We see that $0.5<\alpha_c<1$ in such cases. 
In the inset we show the same set of plots, but for a 
smaller spin separation corresponding to a higher value of 
$f_D(t_d\w_c)=0.3$. In this case, the ratio $X_C/\Delta$ is large enough that once $\Theta(\Delta_r)$ changes its qualitative behavior (i.e. when $\alpha>0.5$), it never 
crosses the line $\Delta$ and the only solution to Eq.~(\ref{eqndeltarselfwithX}) is $\Delta_r=0$. This confirms the limiting behavior, $\alpha_c\approx0.5$ for $X_C/\Delta_r\gg1$, discussed earlier in reference to 
Eqs.~(\ref{eqnbigXdeltar}) and~(\ref{eqnA_Bapprox2}).

\begin{figure}[!t]
\begin{center}
\includegraphics[width=0.45\textwidth]{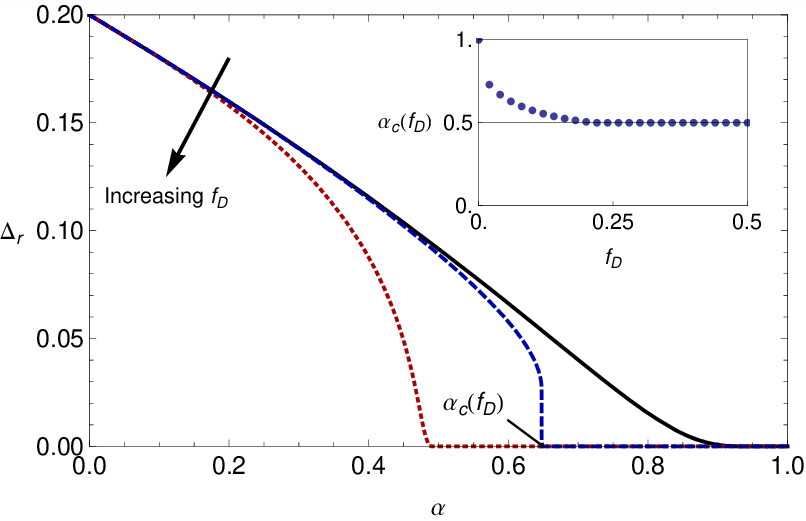}
\caption{(Color online) Main: Numerically evaluated renormalized tunneling strength as a function of the system bath coupling strength for various spin separations, captured by the value of the function $f_D$ measuring the bath correlations. The 
solid black curve corresponds to infinitely separated spins (no Ising interaction), $f_D=0$. The dashed blue curve corresponds to an intermediate spin separation (or Ising interaction), $f_D=0.05$. The dotted 
red curve corresponds to small spin separation (or large Ising interaction), $f_D=0.3$. Inset: Here we show how the critical coupling strength varies with the spin separation. For all curves, $\Delta/\omega_c=0.2$.}
\label{figdeltarwithX}
\end{center}
\end{figure}

Knowing how we expect $\Delta_r$ to behave in certain limits we may now numerically solve Eq.~(\ref{eqndeltarselfwithX}) to find the renormalised tunneling strength as a function of $\alpha$, 
for various spin separations characterized by the function $f_D(t_d\w_c)$ measuring the bath correlations. We shall restrict ourselves here to the two and three dimensional cases, and the results can be seen in Fig.~\ref{figdeltarwithX}. The solid black curve shows the renormalized tunneling for infinitely separated spins, i.e. $f_D=0$ (no bath correlations). As expected, in this regime of zero Ising strength, $\Delta_r\rightarrow 0$ as the system-bath coupling strength 
$\alpha\rightarrow 1$, precisely as in the single spin case. The dashed blue curve shows the variation of the renormalized tunneling with $\alpha$ for a spin separation corresponding to $f_D=0.05$. For this intermediate separation (or, equivalently, Ising strength) we see that $\Delta_r$ discontinuously approaches zero as $\alpha$ reaches a critical value somewhere between $0.5$ and $1.0$ ($\alpha_c\approx0.65$ for the values of $\Delta/\w_c$ and $f_D$ used here). This agrees with the intuition we gained previously from Fig.~\ref{figtheta}. The red dotted curve corresponds to a small spin separation giving $f_D=0.3$ (or large Ising strength). Here, $\Delta_r\rightarrow0$ continuously as $\alpha\rightarrow0.5$, again in agreement with our analysis of Fig.~\ref{figtheta}. In the inset of Fig.~\ref{figdeltarwithX}, we show explicitly how the critical coupling strength depends on the qubit separation. As expected, for large spin separations ($f_D\rightarrow 0$) $\alpha_c$ tends to $1$, while as the spins are brought closer together and $f_D$ increases, $\alpha_c$ approaches its minimum value of $0.5$. 

The last piece of information needed to complete our picture is the value of $f_D$, say $f_{D0}$, after which the crossover always occurs around $\alpha_c=0.5$ (i.e. beyond $f_{D0}$ the bath correlations are large and $X_C$ quickly dominates with increasing $\alpha$). Finding where the minimum of $\Theta(\Delta_r)$ just crosses the line $\Delta$ yields the simple result
\beq
f_{D0}=2 \e \bigg(\frac{\Delta}{\w_c}\bigg)^2.
\label{eqnfD0}
\eeq
For $\Delta/\omega_c=0.2$, we get $f_{D0}\approx0.22$, in agreement with Fig.~\ref{figdeltarwithX}. From this expression it can be seen that the further a given system 
lies within the scaling limit ($\Delta/\w_c \ll1$), the larger the range of spin separations which differ from the single-spin case ($\alpha_c\approx 1$).

\subsection{Section summary}

We conclude this section with a brief summary. For distantly separated spins or negligible Ising strengths, the delocalized-localized crossover corresponds to the well studied single spin-boson model~\cite{chin06, winter09, lehur08, chen08}. The critical coupling strength 
after which the tunneling element is renormalised to zero is predicted to be $\alpha_c=1$, for $\Delta/\omega_c\ll1$. As the spins are brought closer together within their common bath, they become coupled via an Ising-like interaction. This causes the crossover region to drop from $\alpha_c=1$, as can be seen by tracing from 
left to right in the inset of Fig.~\ref{figdeltarwithX}. At a certain spin separation, the function 
scaling the Ising strength, $f_D(t_d\w_c)$, reaches a special value, $f_{D0}$, given by Eq.~(\ref{eqnfD0}). 
For this spin separation, and all smaller separations, the crossover is predicted to occur around $\alpha_c=0.5$.

\section{Full variational treatment}
\label{secexactvariationalcalculation}

\subsection{Free energy minimization and self-consistent equations}

The results presented in the previous section were obtained by approximating the induced Ising strength, $X$, by a value $X_C$, through the replacement $f_{\bf k}\rightarrow g_{\bf k}$.
This significantly simplified the 
task of finding the set of variational parameters $\{f_{\bf k}\}$, which then allowed us to determine the renormalised tunneling strength in a straightforward manner. To go beyond this approximation, we shall now perform the variational calculation making no such simplification, and hence use the full $f_{\bf k}$-dependent Ising strength given in Eq.~(\ref{eqnXdefinition}).

As before, we calculate the free energy associated with the Hamiltonian $\tilde{H}=\tilde{H}_0+\tilde{H}_I$, given by 
Eqs.~(\ref{eqntildeH_0}), (\ref{eqntildeH_z}) and (\ref{eqntildeH_pm}). This leads to an expression for $A_B$ identical to Eq.~(\ref{eqnfreeenergywithX}), but with $X_C$ replaced by $X$: 
\beq
\begin{split}
A_B\approx &\, 2\sum_{\bf k}\w_{\bf k}^{-1}f_{\bf k}(f_{\bf k}-2g_{\bf k})\\
&-\beta^{-1}\txt{ln}\left[2\bigl(\txt{cosh}(2\beta X)+\txt{cosh}(\beta E)\bigr)\right],
\end{split}
\label{eqnfreeenergywithXnotcrude}
\eeq
where $E=\sqrt{4X^2+\Delta_r^2}$. Minimization with respect to the variational parameters $\{f_{\bf k}\}$ gives us the zero-temperature condition
\beq
f_{\bf k}=g_{\bf k} \bigg(\frac{E+2 X \cos\bigl({\bf k}\cdot({\bf r}_1-{\bf r}_2)\bigr)}{E+2X \cos\bigl({\bf k}\cdot({\bf r}_1-{\bf r}_2)\bigr)+\Delta_r^2/\w_k}\bigg),
\label{eqnfkminwithXandd}
\eeq
which is consistent with our assumption $f_{\bf{k}}=f_{-\bf{k}}$, used with reference to 
Eq.~(\ref{eqn_s_commutator}) in our derivation of the transformed Hamiltonian $\tilde{H}$. 
We proceed by inserting Eq.~(\ref{eqnfkminwithXandd}) into our expressions for the renormalized tunneling strength, 
Eq.~(\ref{eqndeltardefinition}), and the full Ising strength, Eq.~(\ref{eqnXdefinition}). For simplicity, we now restrict our discussion to system-bath coupling in three dimensions, in which we may write ${\bf k}\cdot({\bf r}_1-{\bf r}_2)=\w t_d \cos(\theta)$, where $\theta$ is a polar angle 
in ${\bf k}$-space over which we must integrate. We then obtain the following two equations which we must simultaneously solve self-consistently:
\beq
\Delta_r=\Delta\exp{\left[-\frac{\alpha}{2}\int_{-1}^{+1}dx\int_0^{\w_c}\w^{-1}G^2(\w,x)d\w\right]},
\label{eqndeltarselfint}
\eeq
and
\beq
X=\frac{\alpha}{4}\int_{-1}^{+1}dx\int_0^{\w_c}G(\w,x)\bigl(2-G(\w,x)\bigr)\cos(\w x t_d)d\w,
\label{eqnXself}
\eeq
where $x=\cos(\theta)$, and we have defined the function
\beq
G(\w,x)=\bigg(\frac{E+2 X \cos(\w x t_d)}{E+2X \cos(\w x t_d)+\Delta_r^2/\w}\bigg).
\eeq

\begin{figure}[!t]
\begin{center}
\includegraphics[width=0.45\textwidth]{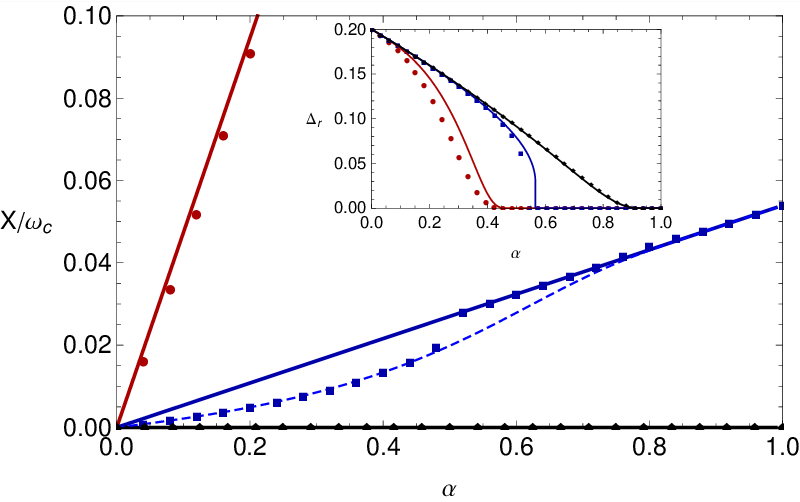}
\caption{(Color online) Main: Comparison of the numerically calculated Ising strength (points) and the crude Ising strength (solid lines) as a function of $\alpha$, with $\Delta/\omega_c=0.2$. This is done for $t_d\w_c=1$ (red line, circular markers), $t_d\w_c=15$ (blue line, square markers) and in the limit 
$t_d\w_c \rightarrow \infty$ (black line, triangular markers, close to the $x$-axis). After the system-bath coupling strength $\alpha=\alpha_c$, $\Delta_r\rightarrow0$ (see inset) and the crude Ising strength matches the full value. The dashed line shows the small Ising strength approximation of Eq.~(\ref{eqn_approx_x}). Inset: Renormalized tunneling strength calculated using the full Ising strength (points) and crude Ising approximation (lines) as a function of $\alpha$, for the same parameters as the main figure.}
\label{fignumeric_X}
\end{center}
\end{figure}

\subsection{Comparison of full and crude Ising strengths}
\label{isingcomp}

Extracting useful analytic expressions from Eqs.~(\ref{eqndeltarselfint}) and (\ref{eqnXself}) is not easily achieved. However, we note that the values $\Delta_r=0$ and $X=X_C$ solve these equations exactly. 
That is, in the localized regime, where the ground state becomes completely dominated by the Ising term, the Ising strength is given by its crude value. 
This tallies with our earlier assertion that the crude Ising approximation is essentially a strong system-bath coupling approximation on the induced interaction strength. Let us also consider the regime in which the spins are distantly separated. 
On physical grounds, we expect that $X\rightarrow 0$ as $|{\bf r}_1-{\bf r}_2|\rightarrow\infty$, since it seems 
inappropriate that the bath could mediate an interaction between spins separated by a large distance 
(certainly, we know that $X_C\rightarrow 0$ as the spin separation is increased to infinity). This can be seen in the present case by making the assumption that 
for large $t_d$ (i.e. large spin separation), $X$ will be small (which we shall justify numerically in the following) and expand the integrand of Eq.~(\ref{eqnXself}) to second order in $X$. 
Having done so, the integrations with respect to $x$ and $\w$ can be performed analytically, leaving a quadratic equation for $X$ which we write as 
\beq
0=h_0(t_d)+X (h_1(t_d)-1)+X^2 h_2(t_d),
\label{eqnquadraticX}
\eeq
where $h_0$, $h_1$, and $h_2$ are cumbersome expressions (proportional to $\alpha$) which we shall not give here. Taking the limit $t_d\rightarrow \infty$, we find that $h_0\rightarrow 0$ and $h_1\rightarrow \alpha\w_c^2/2(\Delta_r+\w_c)^2$. Applying the same limit to $h_2$ is less straightforward, although it is 
easy to see graphically that $h_2\rightarrow0$ as $t_d\rightarrow\infty$. Hence, as expected, we have confirmed that $X\rightarrow 0$ as $t_d \rightarrow \infty$. Further, when $X=0$, the self-consistent equation for the renormalised 
tunneling strength, Eq.~(\ref{eqndeltarselfint}), reduces to that for a single spin given by Eq.~(\ref{eqndeltarself}).

\begin{figure}[!t]
\begin{center}
\includegraphics[width=0.42\textwidth]{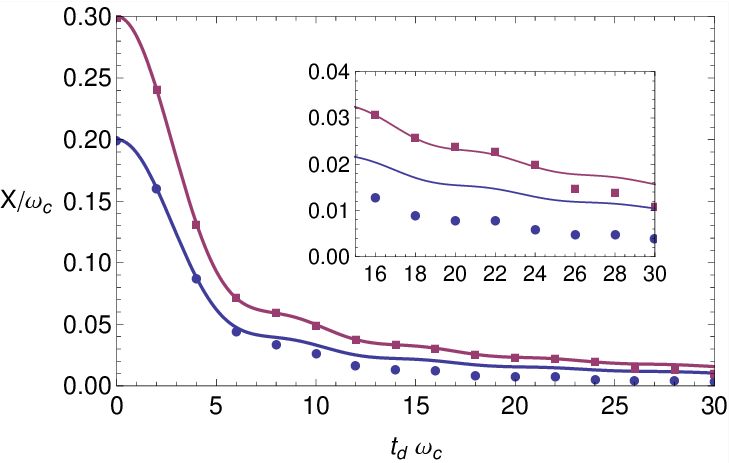}
\caption{(Color online) Main: Induced Ising strength as a function of the scaled spin separation, $t_d\w_c$, for two values of the system-bath coupling strength, $\alpha=0.4$ (blue circular markers) and $\alpha=0.6$ (red square markers). The markers indicate values calculated numerically from 
Eqs.~(\ref{eqndeltarselfint}) and (\ref{eqnXself}) and the solid lines represent $X_C$ values calculated using Eq.~(\ref{eqncrudeXallD}). Inset: Magnification of the lower right corner, revealing how the discrepancy between $X$ and $X_C$ increases when the Ising strength is small enough such that $\Delta_r \neq 0$.}
\label{figXtd}
\end{center}
\end{figure}

When neither the spin separation nor the system-bath coupling strength are large enough such that the above arguments apply, we must solve the self-consistent equations by numerical iteration. Solutions found in this way are shown in Fig.~\ref{fignumeric_X}, where the plot points are 
calculated iteratively from Eqs.~(\ref{eqndeltarselfint}) and (\ref{eqnXself}), and the solid lines calculated using the crude Ising approximation of the previous
section. Red circular points correspond here to a small spin separation, $t_d\w_c=1$, blue squares to an intermediate separation, $t_d\w_c=15$, and black triangles to the limit $t_d\rightarrow \infty$ (for which there is no discrepancy between the full and crude Ising strengths). From the main part of the figure we can see that the crude value of the Ising strength generally gives a reasonably good approximation to the full expression. As the system-bath coupling strength is increased, there comes a point at which the tunneling strength becomes entirely suppressed, $\Delta_r \rightarrow 0$ (see figure inset), in which case Eq.~(\ref{eqnXself}) for $X$ reduces to the simpler form of Eq.~(\ref{eqncrudeXallD}). Hence, in the localized regime $X=X_C$, as expected. 
From the inset of Fig.~\ref{fignumeric_X} we see that the behaviour of the renormalized tunneling strength is well approximated across a range of different parameter regimes by replacing $X$ by $X_C$ in the self-consistent equations. Hence, our analysis of the localization crossover in the two-impurity spin-boson model given in the previous section is expected to hold true, even when the full bath-induced Ising form is used. 

In order in reproduce the behaviour of $X$ for small values of $\alpha$ and moderate spin separations, where it differs most markedly from $X_C$ in Fig.~\ref{fignumeric_X}, we can expand the solution to Eq.~(\ref{eqnquadraticX}) to first order in $\alpha$. In doing so, we find $X\approx h_0(t_d)$, with 
\beq
\begin{split}
&h_0(t_d)=-\frac{\alpha \Delta_r}{2 t_d \mu} \sin(t_d \w_c)\\
&+\frac{\alpha}{2 t_d}(\txt{Ci}(t_d\mu)-\txt{Ci}(t_d \Delta_r))(t_d\Delta_r \cos(t_d\Delta_r)-\sin(t_d\Delta_r))\\
&+\frac{\alpha}{2 t_d}(\txt{Si}(t_d\mu)-\txt{Si}(t_d \Delta_r))(t_d\Delta_r \sin(t_d\Delta_r)+\cos(t_d\Delta_r)),
\end{split}
\label{eqn_approx_x}
\eeq
where $\txt{Ci}(x)=-\int_x^{\infty}\cos(t)/t \txt{d}t$ is the cosine integral function, and we have made the substitution $\mu=\Delta_r+\w_c$. 
The dashed curve in Fig.~\ref{fignumeric_X} shows this function plotted for $t_d \w_c=15$, where we also approximate the renormalised tunneling strength as $\Delta_r\approx \Delta(\Delta/\w_c)^{\alpha/(1-\alpha)}$.

Lastly, in Fig.~\ref{figXtd} we plot a comparison of the behaviour of $X$ and $X_C$ with varying (scaled) spin separation, $t_d \w_c$. Recall that when $\Delta_r=0$, $X=X_C$, as can be seen in the majority of the plot for $\alpha=0.6$. For this value of the system-bath coupling, $\alpha>\alpha_c$ over almost the full range of separations considered, and the tunneling is consequenctly renormalized to zero for most values of $t_d\omega_c$ too. As the spin separation is increased, $X$ decreases, and there comes a point at which $\Delta_r \neq 0$ ($t_d\omega_c\approx26$). 
Here, we begin to see deviations of $X$ from $X_C$. When 
$\alpha=0.4$ the system is always in the delocalized regime ($\Delta_r \neq 0$) and we therefore see deviations of $X$ from $X_C$ for all spin separations. 

\section{Variational ground state}
\label{variationalgs}

\subsection{Two-impurity Hamiltonian in the displaced oscillator basis}

In the preceding sections, we have used a variational treatment to establish how both the renormalised tunneling strength and bath-induced Ising interaction vary as a function of system-bath coupling strength and spin-separation in the two-impurity spin-boson model. We shall now use this information 
to explore the interplay of these two quantities in determining how 
the form of the ground state of the system changes in different parameter regimes. From this, we shall identify a physical indicator of the delocalized to 
localized crossover in the dissipative two-spin system. 

To obtain the variational ground state we generalize the procedure given in 
Section~\ref{secsinglespinbosonmodel} to two spins. We write the total Hamiltonian [Eq.~(\ref{2spinboson})] in a displaced oscillator basis, this time defined by the four states 
$\{\ket{B_{--}}\ket{00},\ket{B_{-+}}\ket{01},\ket{B_{+-}}\ket{10},\ket{B_{++}}\ket{11}\}$, with
\beq
\ket{B_{\pm\pm}}=\prod_{\bf k} D(\pm\alpha_{\bf k} \e^{i{\bf k}\cdot{\bf r}_1})\prod_{\bf k} D(\pm\alpha_{\bf k} \e^{i{\bf k}\cdot{\bf r}_2})\ket{B_0},
\eeq
\noindent and
\beq
\ket{B_{\pm\mp}}=\prod_{\bf k} D(\pm\alpha_{\bf k} \e^{i{\bf k}\cdot{\bf r}_1})\prod_{\bf k} D(\mp\alpha_{\bf k} \e^{i{\bf k}\cdot{\bf r}_2})\ket{B_0},
\eeq
where once again $\alpha_{\bf k}=f_{\bf k}/\w_{\bf k}$ and $\ket{B_0}$ is the state of the bath for vanishing system-bath coupling. In this basis, the two-impurity spin-boson Hamiltonian becomes
\beq
H=-\frac{\Delta_r}{2}(\tilde{\sigma}_x^1+\tilde{\sigma}_x^2)-2X\tilde{\sigma}_z^1\tilde{\sigma}_z^2+2R,
\eeq
where the zero temperature limit has been taken, and $R$, $\Delta_r$, and $X$ are defined in Eqs.~(\ref{eqn:R}),~(\ref{eqndeltardefinition}), and~(\ref{eqnXdefinition}), respectively. Diagonalizing this Hamiltonian gives a ground 
state energy of $\Lambda_0=2R-E$, and corresponding ground state
\beq
\begin{split}
\ket{\Phi_0}=N_0\Bigl(&\ket{B_{--}}\ket{00}+\ket{B_{++}}\ket{11}\\
&-\xi\bigl(\ket{B_{+-}}\ket{10}+\ket{B_{-+}}\ket{01}\bigr)\Bigr),
\label{eqntotalstate}
\end{split}
\eeq
where $N_0=(2(1+\xi^2))^{-1/2}$, $\xi=(2X-E)/\Delta_r$, and $E=\sqrt{4X^2+\Delta_r^2}$ as before. 
Minimizing $\Lambda_0$ with respect to the variational parameters leads to exactly the same condition [Eq.~(\ref{eqnfkminwithXandd})] as derived in 
Section~\ref{secexactvariationalcalculation}. Therefore, we shall make the crude Ising approximation 
to evaluate $X$ and $\Delta_r$, giving all of the required information relating to the variational ground state.

\begin{figure}[!t]
\begin{center}
\includegraphics[width=0.45\textwidth]{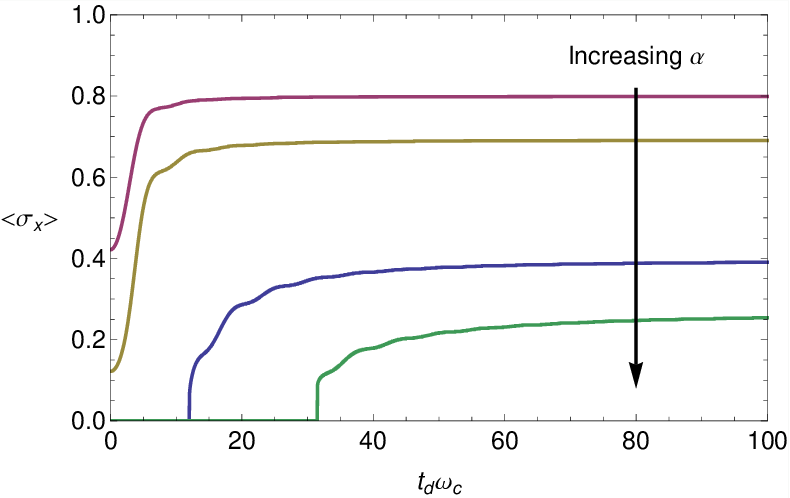}
\caption{(Color online) Expectation value of $\sigma_x^1$ (or $\sigma_x^2$) as a function of the scaled spin separation for different values of the system bath coupling strength $\alpha$ (=0.20, 0.30, 0.55, 0.65 ordered as indicated). 
For $\alpha=0.55$ and $\alpha=0.65$ we see that at a particular spin separation there emerges a non-zero 
expectation value, signifying the crossover from localization to delocalization.}
\label{figsigmax}
\end{center}
\end{figure}

\subsection{Experimental signatures of localization-delocalization crossover}

To show how evidence for the localization crossover might be observed experimentally, 
in Fig.~\ref{figsigmax} we plot the ground-state expectation value of the {\it single-spin} operator $\sigma_x^1$ (or equivalently $\sigma_x^2$), 
$\av{\sigma_x^1}=\bra{\Phi_0}\sigma_x^1\ket{\Phi_0}=-2\xi\av{B}/(1+\xi^2)$, as a function of the scaled spin separation 
for various values of the system-bath coupling strength. For small values of $\alpha$ ($\alpha=0.2,0.3$) the tunneling 
element is renormalised to a finite value (delocalized regime) and $\av{\sigma_x^1}$ is predominantly determined by the relative size of the bare tunneling element to the Ising strength, saturating at a value $\av{\sigma_x^1}\approx\av{B}$ at large spin separations (small $X$). There is no qualitative change in the ground-state form as the relative size of $\Delta_r$ and $X$ varies, in this case through increasing the spin separation.  

\begin{figure}[!t]
\begin{center}
\includegraphics[width=0.45\textwidth]{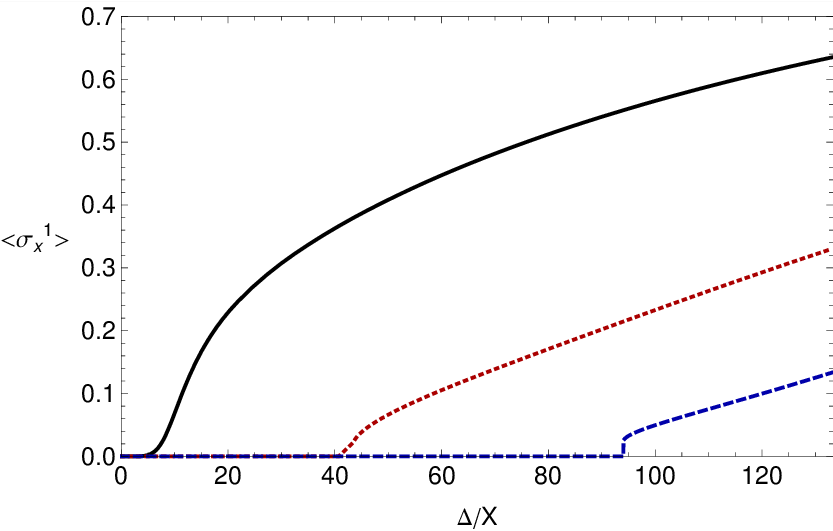}
\caption{(Color online) Expectation value of $\sigma_x^1$ (or $\sigma_x^2$) as a function of $\Delta$ (measured in units of $X$) for $\alpha=0.35$ (black solid line), $\alpha=0.6$ (red dotted line), and $\alpha=0.75$ (blue dashed line). 
For these plots the induced Ising strength was kept at $X=0.0015\w_c$ for each $\alpha$, with $\w_c=1$.}
\label{figsigx_vs_delta}
\end{center}
\end{figure}

For larger values of $\alpha$, lying between $0.5$ and $1$, the Ising strength at small spin separations is large enough such that the renormalized tunneling strength is completely suppressed, and $\av{\sigma_x^1}\rightarrow 0$ (localized regime). As the spin separation increases, the Ising strength decreases, and there comes a point at which $X$ is small enough such that $\av{\sigma_x^1}$ can now take on non-zero values (delocalized) {\it for the same value of $\alpha$}. Therefore, if it is possible to engineer a pair of Ising-coupled spins for which the Ising strength can be varied, and $0.5 <\alpha<1$, the crossover region should be identifiable by the emergence of a non-zero value for $\av{\sigma_x^1}$ (or $\av{\sigma_x^2}$) as the Ising interaction is decreased.

It is also possible to observe the crossover behaviour without the need for varying the Ising strength, by instead altering the bare tunneling frequency due to the applied field.
In Fig.~\ref{figsigx_vs_delta} we again plot $\av{\sigma_x^1}$ but this time as a function of the bare tunneling strength with fixed bath mediated Ising strength $X$. For $\alpha=0.35$ we expect no 
crossover in ground state behaviour and we see $\av{\sigma_x^1}\rightarrow 0$ only as $\Delta \rightarrow 0$. For the curves corresponding to $\alpha>0.5$, when $\Delta/X$ is small we are in the regime in which Eq.~(\ref{eqnbigXdeltar}) is valid. 
As such, $\alpha_c=0.5$ and we see $\av{\sigma_x^1}=0$. As the ratio $\Delta/X$ is increased, we eventually move into a regime in which Eq.~(\ref{eqn_standard_deltar}) is valid and $\alpha_c\rightarrow 1$. For $\alpha=0.6$ and $\alpha=0.75$ 
we must therefore enter the delocalized regime as $\Delta/X$ increases, and $\av{\sigma_x^1}$ thus begins to take on non-zero values.

\subsection{System-bath entanglement}

Quantum phase transitions are associated with non-analyticity in the entanglement present in the total system-plus-bath state~\cite{wu04, lambert05, kopp07, rieper2010}. 
Although the variational treatment may not identify a true quantum phase transition, it is expected that the change in ground-state properties that are identified 
will have a manifestation in the entanglement~\cite{lehur08}. 
Since, within the variational approach, the total state [Eq.~(\ref{eqntotalstate})] is a pure state, we can investigate such behaviour in our model simply by tracing 
out the bath degrees of freedom and calculating the von Nuemann entropy of the two-spin state. This will give a measure of the degree to which 
the spins are entangled with the bath~\cite{n+c}. We define the reduced two-spin state as $\rho={\rm tr_B}(\ketbra{\Phi_0}{\Phi_0})$, where ${\rm tr_B}$ denotes a trace 
over the bath degrees of freedom. The von Neumann entropy is then defined as
\beq
S=-\rho \txt{ln}(\rho)=-\sum_{i=1}^{4}\tau_i\txt{ln}(\tau_i),
\eeq
where the $\tau_i$ are the four eigenvalues of $\rho$~\cite{n+c}.

\begin{figure}[!t]
\begin{center}
\includegraphics[width=0.45\textwidth]{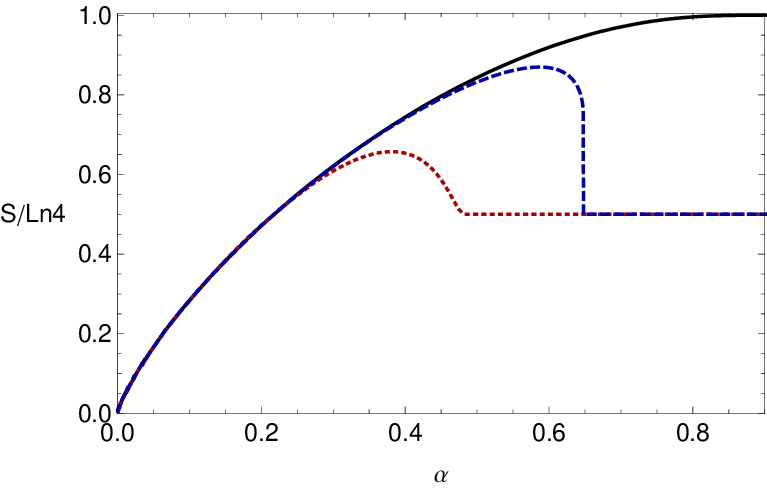}
\caption{(Color online) Normalized von Neumann entropy of the two-spin system as a function of $\alpha$ for $f_D=0$ (black solid line), 
$f_D=0.05$, (blue dashed line) and $f_D=0.3$ (red dotted line), with $\Delta/\w_c=0.2$.}
\label{figentropy}
\end{center}
\end{figure}

In Fig.~\ref{figentropy} we plot the entropy as defined above (normalised by its maximum possible value)  for three different spin separations, 
corresponding to $f_D=0$ (black solid line), $f_D=0.05$ (blue dashed line) and $f_D=0.3$ (red dotted line). 
For $f_D=0$ the situation is identical to the single spin case. As $\alpha$ is increased, the extent to which the spins and the bath interact increases and 
their state becomes ever more entangled. For the curves corresponding to $f_D=0.05$ and $f_D=0.3$ we see a similar situation for small values of 
$\alpha$. However, for moderate values of $\alpha$ we see that the entanglement reaches a maximum and then begins to fall. This corresponds 
to the onset of the crossover between delocalization and localization in the ground state. At the critical values of $\alpha$ for these spin separations 
($\alpha_c=0.65$ and $\alpha_c=0.5$, respectively, for these parameters), the entanglement sharply drops to a value of $0.5$ as 
$\Delta_r\rightarrow 0$. For a single spin $S=0$ in its localized regime~\cite{lehur08}. In the present case we find 
$S/\ln4=0.5$ since there is nothing in our model to lift the degeneracy between the states $\ket{00}$ and $\ket{11}$ 
in the localized regime.

\section{Summary}
\label{secsummary}

We have investigated the delocalized to localized crossover for a pair of two-level systems in a common bosonic bath. Our analysis followed closely that introduced for single spins by Silbey and Harris~\cite{silbey84} which used a variational approach. The crossover region is 
identified by a complete suppression of the tunneling element ($\Delta_r \rightarrow 0$) as the system-bath coupling is increased ($\alpha \rightarrow \alpha_c$). We find an interesting interplay between the magnitude of an environment-induced Ising spin interaction ($X$) and the applied tunneling field ($\Delta$) in determining $\alpha_c$. In particular, our analysis suggests that the presence of the Ising term {\it encourages} the spins to enter the localised regime at a smaller value of the system-bath interaction than in the single spin case. Specifically, only for infinitely separated spins do we recover $\alpha_c=1$, as in the single spin-boson model. On reducing the spin separation from infinity, $\alpha_c \rightarrow 0.5$. Interestingly, 
$\alpha_c$ reaches this minimum value at a finite spin separation, and retains this value for all smaller separations. We also obtained the variational ground state, and from this showed that 
a signature of the groud state crossover could be found in the emergence of a finite single-spin expectation value
$\langle \sigma_x \rangle$ as either the spin separation or the ratio of tunneling strength to Ising interaction is increased. The crossover should 
also be evident in the entanglement shared between the system and bath.

\subsection{Acknowledgements}

The authors wish to thank Alex Chin and Janet Anders for useful and interesting discussions. S.B. thanks the Royal Society and Wolfson Foundation. 
D.P.S.M., A.N. and S.B. are supported by the {\sc epsrc}.

\end{document}